\documentclass[pra,twocolumn,showpacs,preprintnumbers,amsmath,amssymb,tightenlines]{revtex4}

\usepackage{graphicx}
\usepackage{dcolumn}
\usepackage{bm}
\usepackage{epsfig}
\usepackage{stmaryrd}

 \newcommand{\pdiff}[2]{\frac{\partial #1}{\partial #2}}
 \newcommand{\pdiffn}[3]{\frac{\partial^{#3} #1}{\partial #2^{#3}}}

\newcommand{\tabspace}{0cm}
\newcommand{\sub}[2]{{#1}_{\mbox{\!\! \scriptsize #2}}}

\def\dl{\tilde}

\def\beq{\begin{equation}}
\def\eeq{\end{equation}}

\def\CR{\nonumber\\[0.15cm]}
\newcommand{\rref}[1]{Ref.~\cite{#1}}
\newcommand{\fref}[1]{Fig.~\ref{#1}}
\newcommand{\frefp}[2]{Fig.~\ref{#1}~(#2)}

\newcommand{\eref}[1]{Eq.~(\ref{#1})}
\newcommand{\esref}[2]{Eqs.~(\ref{#1}) and (\ref{#2})}

\newcommand{\sref}[1]{section~\ref{#1}}

\newcommand{\cref}[1]{chapter~\ref{#1}}

\newcommand{\Cref}[1]{Chapter~\ref{#1}}
\newcommand{\tref}[1]{table~\ref{#1}}

\newcommand{\bref}[1]{(\ref{#1})}

\begin{document}

\title{Limits to the analogue Hawking temperature in a Bose-Einstein condensate}
\author{S. W\"uster and C.~M. Savage}
\affiliation{ARC Centre of Excellence for Quantum-Atom Optics, Australian National University, Canberra ACT 0200, Australia}
\email{craig.savage@anu.edu.au}

\begin{abstract}
Quasi-one dimensional outflow from a dilute gas Bose-Einstein condensate reservoir is a promising system for the creation of analogue Hawking radiation. We use numerical modeling to show that stable sonic horizons exist in such a system under realistic conditions, taking into account the transverse dimensions and  three-body loss. We find that loss limits the analogue Hawking temperatures achievable in the hydrodynamic regime, with sodium condensates allowing the highest temperatures. A condensate of 30,000 atoms, with transverse confinement frequency $\omega_{\perp}=6800\times 2\pi$ Hz, yields horizon temperatures of about $20$ nK over a period of $50$ ms. This is at least four times higher than for other atoms commonly used for Bose-Einstein condensates.
\end{abstract}


\maketitle


\section{Introduction}
 Bose-Einstein condensates (BECs) are a promising physical system with which to realize the analogue gravity program~\cite{visser:review}. This field advocates the observation of the analogues of phenomena such as:  Hawking radiation~\cite{unruh:bholes}, cosmological particle production~\cite{barcelo:cpc} and superradiance~\cite{tracy:superradiance}. The analogies follow from the equivalence between the equations of motion for a scalar quantum field in curved space-time and for the phonons of a flowing BEC in the hydrodynamic regime~\cite{unruh:bholes}. The interest in the analogue program arises because of the lack of prospects for experiments in the gravitational domain.
 
 A {\it sonic horizon} in a BEC is a surface on which the normal component of the condensate flow exceeds the speed of sound.  It is analogous to the {\em event horizon} of a black hole. Sonic horizons are predicted to  Hawking radiate  phonons~\cite{unruh:bholes}.
 A variety of methods to create sonic horizons in BECs have been suggested, most of which generate both black and white hole horizons~\cite{visser:towards,sakagami:hr,garay:prl,garay:pra,us:nozzle}. A white hole horizon is a surface on which the condensate flow drops back below the speed of sound.
Due to the instabilities associated with white holes, methods which do not generate them are attractive, such as outflow from a reservoir~\cite{giovanazzi:horizon}. 
In this paper we study a simplified version of the latter method, under realistic experimental conditions, including  three-dimensions and three-body loss.
 
In general relativity, Hawking radiation~\cite{hawking:hr1,hawking:hr2} is a quantum effect whereby a black hole horizon emits a thermal spectrum of particles. In the analogous sonic horizon case the emission temperature is given, for one-dimensional flow, by~\cite{visser:towards}
\begin{align}
T_{H}&=\frac{\hbar g_{h}}{2\pi k_{B} c},\:\:
\label{hawkingtemp1}
\end{align}

\begin{align}
g_{h}&=\frac{1}{2}\left| \frac{\partial \left(c^{2} - v^{2} \right)}{\partial x} \right|_{x=x_{h}} = c^{2} \left| \frac{\partial M}{\partial x}\right|_{x=x_{h}} .
\label{surfacegrav}
\end{align}
$v$ is the flow speed, $c$ the speed of sound, $M=v/c$ the Mach number and $x_{h}$ denotes the position of the horizon, where $v=c$. 

We find that three-body loss severely limits the Hawking temperature that can be achieved in BECs, independent of the detailed implementation of the sonic horizon. Amongst helium, rubidium, cesium and sodium, the latter allows the highest Hawking temperature, for fixed total atom number loss, because of its relatively low three-body loss rate and high two-body interaction strength. 

This paper is organized as follows. In \sref{methods}, we introduce our notation and formalism, after which we derive a general limit on the analogue Hawking temperature in \sref{horizon_temperature}.  Numerical studies of horizons generated by reservoir outflow are presented in \sref{analogue_outflow}. 

\section{Methodology \label{methods}}
The dynamics of the BEC wavefunction $\phi$, including three-body losses, is described by the Gross-Pitaevskii equation (GPE) \cite{savage:coll}
\begin{align}
i \hbar  \frac{\partial \phi}{\partial t}&=\left(-\frac{\hbar^2}{2m} \mathbf{\nabla}^{2} +V
 +U  |\phi |^{2} -i\hbar\frac{K_{3}}{2}|\phi|^{4} \right)  \phi.
\label{3dgpe}
\end{align}
Here $m$ is the atomic mass, $U $ the interaction related to the scattering length $a_{s}$ by $U =4 \pi \hbar^{2} a_{s}/m$ and $K_{3}$ the three-body loss coefficient. $V$ denotes an external potential. Using $\phi=\sqrt{n}\exp{(i \vartheta)}$, we can extract the atom number density $n$ and velocity $\mathbf{v}=\hbar\mathbf{\nabla}\vartheta/m$. \eref{3dgpe} can be rewritten in terms of $n$ and $\mathbf{v}$, resulting in continuity and Bernoulli equations for the BEC. For now we assume $K_{3}=0$ and that the condensate does not vary on length scales shorter than the healing length $\xi=1/\sqrt{8\pi n a_{s}}=\hbar/(\sqrt{2}mc)$. We also assume that the system can be modelled in one dimension: for example, flow in a narrow confining potential with longitudinal coordinate $x$ and effective cross-section $A(x)$.
We can then write the approximate equations \cite{giovanazzi:horizon}:
\begin{align}
J&=n(x) v(x) A(x),
\label{continuityeqn2}
\\
\mu&=\frac{1}{2}m \mathbf{v}^{2}(x) + U n(x) + V(x).
\label{hydrodyn2}
\end{align}
Here $\mu$ is the chemical potential, $J$ the particle flux and $v$ the flow speed $v=|\mathbf{v}|$. The local speed of sound $c=\sqrt{U n/m}$. Without assuming the absence of short length scale variations in the condensate variables, the so called quantum pressure $Q(x)=-\frac{\hbar^{2}}{2m}\frac{\nabla^{2}\sqrt{n}}{\sqrt{n}}$ would contribute to the rhs.~of \eref{hydrodyn2}.
Combining Eqs. (\ref{continuityeqn2}) and (\ref{hydrodyn2}), we obtain~\cite{giovanazzi:horizon}:
\begin{align}
&q(x)M(x)^{2/3} -\frac{1}{2}M^{2}(x)=1, 
\label{flowequation}
\\
&q(x)=\frac{\mu-V(x)}{m s(x)^{2}}, \:\:\:\:\:\:s(x)=\left(\frac{J U}{A(x)m}\right)^{1/3}.
\label{qparam}
\end{align}
The function $q(x)$ contains all information about the trapping configuration. The quantity $s(x)$ equals the speed of sound at the horizon $s(x_{h})=c(x_{h})$, see \eref{continuityeqn2}. Differentiating with respect to $x$, solving for the partial derivative with respect to position $M'(x)$, and applying L'H{\^o}spital's rule, we obtain for the horizon temperature,
\begin{align}
T_{H}&=\frac{\hbar c}{2\pi k_{B}}|M'|=\frac{\hbar c}{2\pi k_{B}}\frac{\sqrt{3q''}}{2} ,
\label{machderiv}
\end{align}
where all variables are evaluated at the horizon.
To reach this simple form we used $M(x_{h})=1$, $q(x_{h})=3/2$, see \eref{flowequation}, and $q'(x_{h})=0$~\cite{giovanazzi:horizon}. The latter implies the regularity condition \cite{liberati:unexpected}
\begin{align}
\left(c^{2}A'/A -V'/m \right) \Large|_{x=x_{h}}&=0 .
\label{regularity}
\end{align}
From \eref{qparam} we get
\begin{align}
q''&=\frac{1}{s^{2}}\left(\frac{\mu-V}{m}(\log{A})''-\frac{V''}{m} -\frac{4}{3}(\log{A})'\frac{V'}{m}\right).
\label{qdd}
\end{align}
While analytic expressions for the horizon temperature in the cases of constant potential $V$, or constant area $A$, have been previously obtained \cite{visser:towards, giovanazzi:horizon}, \esref{machderiv}{qdd} are valid where both vary.

In several simulations we use the one dimensional Gross-Pitaevskii equation. For these cases, we assume that the BEC is harmonically confined in the transverse directions, with a large trapping frequency $\omega_{\perp}$. We can then perform a dimensional reduction of \eref{3dgpe}, assuming a fixed Gaussian transverse wave function, $\phi(x,r,t)=\exp{(-r^2/(2 a_{osc}^2 ))}/\sqrt{\pi a_{osc}^2} \ \psi(x)$, where $r$ is the transverse radial coordinate and $a_{osc}=\sqrt{\hbar/(m \omega_{\perp})}$. Then, we obtain an effective one-dimensional interaction strength $U_{1D}=U /(2\pi a_{osc}^{2})$ and loss coefficient $K_{3,1D}=K_{3}/(3\pi^{2} a_{osc}^{4})$.

All numerical solutions were obtained using dimensionless units (tilded in the following), with $\tilde{\phi}= \phi \: a_{osc}^{3/2}$, $\tilde{t}=t \omega_{\perp}$, $\tilde{\mathbf{x}}=\mathbf{x}/a_{osc}$. 
Then $\tilde{U} =4\pi a_{s} / a_{osc}$, $\tilde{U}_{1D}=\tilde{U} /(2\pi)$,  $\tilde{K}_{3}=K_3/ (\omega_{\perp}a_{osc}^6 )$ and $\tilde{K}_{3,1D}=\tilde{K}_{3}/(3\pi^{2})$. The explicit form of the one-dimensional GPE after conversion to dimensionless units is
\begin{align}
i  \frac{\partial \tilde{\psi}}{\partial \tilde{t}}&=\left(-\frac{1}{2} \pdiffn{}{\tilde{x}}{2} +\tilde{V}
 +\tilde{U}_{1D} |\tilde{\psi} |^{2} -i\frac{\sub{\tilde{K}}{3,1D}}{2}|\tilde{\psi}|^{4} \right)  \tilde{\psi}.
\label{1dgpe}
\end{align}
Note that the 1D density $\tilde{n}_{1D}= |\tilde{\psi} |^{2}$ and 3D density $\tilde{n}= |\tilde{\phi} |^{2}$ are connected by $\tilde{n}_{1D}=\pi \tilde{n}$. We define the dimensionless Hawking temperature 
\begin{align}
\tilde{T}_{H}= \frac{c }{2\pi}\left| M' \right|_{x=x_{h}} ,
\label{dimless_th}
\end{align}
satisfying $T_{H}=\omega_{\perp} \hbar \tilde{T}_{H}/k_{B}=\omega_{\perp}7.6 \times 10^{-12}\;\tilde{T}_{H}$. When presenting our results using dimensionless units, we always state a concrete atomic system to which they pertain. The purpose of the dimensionless units is then to allow simple conversion of the results between different atomic species, which is possible for our results without three-body losses.

As our focus is on experimental realism, we base our results on numerical solutions of the full GPE. However, we make use of the approximations Eqs.~\bref{continuityeqn2}-\bref{qdd} to guide these numerical simulations. In particular, we can use \eref{machderiv} to estimate the analogue Hawking temperature from a given potential, and the regularity condition \bref{regularity} to predict the location of the sonic horizon.

\section{Limits on the Analogue Hawking Temperature\label{horizon_temperature}}

The analogy between quantum fields in curved space-time and the excitations in a BEC requires both  the condensate and excitations to be in the hydrodynamic regime \cite{barcelo:diffmetric}. The former requires variations in the density and speed to occur only on length scales much larger than the healing length $\xi$, and the latter confines our analysis to phononic modes, i.e.~those whose wave numbers $q$ fulfill $q\xi\ll1$. 

Here, we show that for a condensate horizon in the hydrodynamic regime, the achievable analogue Hawking temperature is limited by three-body loss processes. We formalize the condition that the flow has to remain in the hydrodynamic regime by demanding
\begin{align}
\left| \pdiff{M}{x} \right|_{x=x_h}\lesssim \frac{1}{D\xi},
\label{hydrodynflow_condition}
\end{align}
for $D\gg 1$, where $D$ is a constant chosen to specify the flow speed gradient we are judging to be acceptable. If the Mach number profile is roughly linear near the horizon, we are requiring that $M$ varies at most by the fraction $1/D$ within the space of one healing length $\xi$. Inserting \eref{hydrodynflow_condition} and the expression for $\xi$ into \eref{hawkingtemp1} we obtain
\begin{align}
T_{H}&\lesssim \frac{mc^{2}}{\sqrt{2}\pi k_{B} D}=\frac{U n}{\sqrt{2}\pi k_{B} D}.
\label{temperaturelimit1}
\end{align}
All quantities are evaluated at the horizon. Hence high analogue Hawking temperatures require dense condensates. However, losses limit the achievable densities. 

To determine a quantitative limit on the analogue Hawking temperature, we investigate the effect of the loss in detail. We consider a flow that is harmonically confined with frequency $\omega_{\perp}$ in the transverse directions, and is in either of two regimes: We call the condensate quasi-one dimensional (Q1D) when $\mu\ll \hbar \omega_{\perp}$. In this case transverse excitations are ``frozen'', and we can factor the BEC wave function into longitudinal and transverse parts; the latter remaining in the oscillator ground state \cite{pearl:q1d}. 
The velocity is then exclusively in the longitudinal $x$ direction. Our second regime is $\mu\gg \hbar \omega_{\perp}$, which we call ``transverse Thomas-Fermi'' (TTF). Then we can consider the flow in the Thomas-Fermi approximation \cite{book:pethik} and the velocity can in general have a transverse component. 
We assume the density profile
\begin{align}
\sub{n}{q1d}(x,r,t)&=\sub{n}{0}(x,t)\exp{\left(-\frac{r^{2}}{\sigma^{2}_{r}}\right)}
\label{transverse_q1d}
\end{align}
in the Q1D case. Note that, in this section only, the normalisation of the transverse wave function differs from that presented in \sref{methods} to give $\sub{n}{0}(x,t)$ the meaning of the actual time dependent peak density. 

In the TTF regime we assume
\begin{align}
\sub{n}{ttf}(x,r,t)&=\sub{n}{0}(x,t)-\frac{1}{2}\frac{m \omega^{2}_{r}r^{2}}{U},
\label{transverse_ttf}
\end{align}
with $\sub{n}{ttf}(x,r,t)=0$, where \eref{transverse_ttf} would be negative. 
 
If we assume that the transverse structure of the condensate is always well described by either of the profiles defined above, we can integrate out the transverse coordinates. In this case, the effective loss-equation for the peak density is
\begin{align}
\pdiff{\sub{n}{0}(x,t)}{t}&=-\frac{1}{\alpha_3}K_3\sub{n}{0}(x,t)^{3},
\label{peakdensloss}
\end{align}
with $\alpha_{3}=3$ in the Q1D and $\alpha_{3}=4$ in the TTF regime. 

For short times or weak loss, we can approximate \eref{peakdensloss} by a finite difference equation. We now demand that losses at most reduce the peak density by a fraction $f$ of the initial value within a time interval $\Delta t$. This corresponds to 
\begin{align}
 \frac{K_{3}}{\alpha_{3}}n_{0}^{3}\Delta t &\lesssim f n_{0}.
\label{densitylimit1}
\end{align}
We hence define the maximal allowed peak density under these conditions:
\begin{align}
\sub{n}{max}&=\sqrt{\frac{\alpha_{3} f}{K_{3}\Delta t}}.
\label{peakdens}
\end{align}
Finally we substitute \eref{peakdens} for the density into \eref{temperaturelimit1} and obtain
\begin{align}
T_{H}&\lesssim \frac{U}{\sqrt{2}\pi k_{B} D}\sqrt{\frac{\alpha_{3} f}{K_{3}\Delta t}}.
\label{temperaturelimit2}
\end{align}
This limit on the analogue Hawking temperature is one of our key results.

\subsection{Comparison of BEC-Atom Species}

In the light of the preceding result, we now compare atomic species commonly employed in BEC experiments: $^{4}$He, $^{23}$Na, $^{87}$Rb and $^{137}$Cs. The relevant parameters for these atoms are listed in \tref{atomicparameters}.
\begin{table}[hbt]
\begin{center}
\begin{tabular}{|c|ccc|}
\cline{1-4}
  \Big. atom      & \hspace{\tabspace}$U\times10^{50}$ [Jm$^3$] \hspace{\tabspace}  \hspace{\tabspace}& \hspace{\tabspace} $K_{2}\times10^{20}$ [m$^{3}/$s] & \hspace{\tabspace} $K_{3}\times10^{42}$ [m$^{6}/$s]   \\
   \cline{1-4}
$^{4}$He     & $15.7$    & $2$, \rref{vassen:heliumbec} & $9000$, \rref{vassen:heliumbec} \\
$^{23}$Na     & $1$     & $0.053$, \rref{ketterle:sodk3} & $2.12$, \rref{ketterle:sodk3} \\
$^{87}$Rb     & $0.5$     & $150$, \rref{duerr:rbk2} & $32$, \rref{volz:feshbachs} \\
$^{137}$Cs   & $0.66$  & \mbox{small}  & $130$, \rref{grimm:largea} \\
\cline{1-4}
\end{tabular}
\end{center}
\caption{Parameters for atomic species. $K_2$ is the two-body loss coeffcient. The values for cesium assume a magnetic field of $23$ G, where the ratio of scattering length to three-body loss is most favorable~\cite{grimm:largea}.}
\label{atomicparameters}
\end{table}

According to Eq.~(\ref{temperaturelimit2}) the combination of atomic parameters that determines the limiting Hawking temperature is $U/\sqrt{K_3}$. For $^{4}$He, $^{23}$Na, $^{87}$Rb and $^{137}$Cs this has the respective values: 1.7, 6.9, 0.9, and 0.6, in units of $10^{-30}$ Js$^{1/2}$. The limiting Hawking temperature is four to ten times higher for  $^{23}$Na than for the other species.

The scattering length $a_{s}$, and hence the interaction strength $U$, can be tuned towards higher, more favorable values with the use of Feshbach resonances. However in the large $a_{s}$ regime, the three-body loss rate increases as $a_{s}^{4}$~\cite{grimm:largea}. According to \eref{temperaturelimit2} this rules out an improvement of $T_{H}$ using Feshbach resonances. 

In order to obtain numerical temperatures from \eref{temperaturelimit2} we must specify the maximum Mach number gradient and the maximum acceptable loss rate constraints, which are determined by the parameters $D$ and $f / \Delta t$ respectively.
We choose $D=20$, which places the sonic horizon marginally within the hydrodynamic regime. Note that to use the eikonal approximation for a phonon excitation near the horizon \cite{visser:essential}, the condensate flow should vary little within one wavelength $\lambda=2\pi/q$, where $q$ is the wave number. 
For the shortest wavelength phonons $1/q\sim\xi$. Hence we require small Mach number variations within $\lambda\sim2\pi\xi$ which is barely true for $D=20 \sim 6\pi$. The necessity to consider phononic modes places stronger constraints on the flow variation than the requirement that the flow itself be in the hydrodynamic regime.
\eref{temperaturelimit2} is weakly dependent on the maximum loss rate $f / \Delta t$. We choose to consider a loss of $10\%$ of the peak density ($f=0.1$) within a time $\Delta t=50$ ms, so that $f / \Delta t = 2$ s$^{-1}$. The maximum Hawking temperatures using these criteria are listed in \tref{maxtemperatures1}.
\begin{table}[hbt]
\begin{center}
\begin{tabular}{|c|cccc|}
\cline{1-5}
  \Big. 
  atom      & \hspace{\tabspace}$\sub{n}{max} \times 10^{-19}$ [m$^{-3}$] \hspace{\tabspace}  &  \hspace{\tabspace} $\xi$[$\mu$m] \hspace{\tabspace}  &  $T_H$ [nK] &  $T_H/\sub{T}{crit}$\\
   \cline{1-5}
$^{4}$He     & $3.0$   & $0.42 $    & $3.8$  & $0.10\%$\\
$^{23}$Na     & $194$    & $0.086 $    & $16$  & $0.15\%$\\
$^{87}$Rb     & $50$   & $0.12 $   & $2.2$  & $0.19\%$\\
$^{137}$Cs   & $25$    & $0.12 $   & $1.4$  & $0.30\%$\\
\cline{1-5}
\end{tabular}
\end{center}
\caption{Achievable Hawking temperatures according to \eref{temperaturelimit2}, calculated using $\alpha_{3}=4$, $\Delta t=0.05$ s, $f=0.1$ and $D=20$. Also shown are the condensate densities and associated healing lengths allowed by criterion \eref{peakdens}. The temperatures are also given in relation to the critical temperature $\sub{T}{crit}=\left( n / \zeta(3/2) \right)^{3/2} 2\pi\hbar^{2} / (m k_{B} )$.
\label{maxtemperatures1}}
\end{table}

The values vary by more than an order of magnitude between the atomic species, with sodium exhibiting the highest temperature of 16 nK. We also give $T_{H}$ in relation to the condensation temperature \cite{footnote:tcrit}. For the densities in column one of \tref{maxtemperatures1}, two-body losses are negligible compared to three-body losses for the values of $K_{2}$ in \tref{atomicparameters}.

The temperatures in \tref{maxtemperatures1} compare well with the 15 nK given for sodium in \rref{giovanazzi:horizon}. The values are low in spite of the optimistic values for $\Delta t$, $D$ and $f$ that we have used. By allowing large losses ($f=0.1$) in a short time ($\Delta t=50$ ms), we are assuming quick phonon detection. Slower detection, and moving the horizon safely into the hydrodynamic regime, would require all three values to be adjusted in the direction producing decreased Hawking temperature.  Hence fast phonon detection schemes such as proposed in \rref{schuetzhold:phonondetection} are attractive. Another useful tool would be suppression of inelastic collisions \cite{yurovsky:lessloss}.

An interesting direction for future research is motivated by the low temperatures of \tref{maxtemperatures1}: Determining whether an emission of quanta also takes place in condensates in which $\pdiff{M}{x}$ exceeds the hydrodynamic bound of \eref{hydrodynflow_condition} (or $D\sim1$) and how visible this radiation would be. A candidate for such a condensate flow is that past a gray soliton \cite{us:nozzle}.
As for $D\sim1$ strong deviations from a thermal spectrum are expected outside of the hydrodynamic regime, the use of \eref{hawkingtemp1} for the radiation temperature no longer makes sense there. 

It has also been suggested to realise the analogue Hawking effect in degenerate Fermi gases \cite{giovanazzi:fermions, giovanazzi:fermionslong}, for which three-body recombination would be suppressed.

\section{Reservoir Outflow\label{analogue_outflow}}
Many horizon systems suggested in the literature possess a white hole in addition to the black hole~\cite{visser:towards,sakagami:hr,garay:prl,garay:pra,us:nozzle} . A white hole is the surface where the flow speed returns below the speed of sound.  It has been shown that the condensate develops dynamical instabilities in the presence of a white hole horizon~\cite{garay:prl,leonhardt:instability,garay:bhstab}. While these can be suppressed by appropriate boundary conditions~\cite{garay:bhstab}, in a realistic experiment they are likely to be present. In a range of numerical investigations of sonic horizons in 1D and 2D, we found dynamical instability in all cases containing a white hole. It manifests itself in the repeated creation of gray solitons~\cite{us:nozzle}. 

The reservoir outflow scenario of Giovanazzi {\it et al.}~\cite{giovanazzi:horizon} has a black hole horizon without a white hole horizon.  They reported, using one-dimensional simulations without losses, a setup with dynamically stable horizon formation.
We analyze a simplified version of this model in greater detail, as it is the most promising method for experimentally realizing a stable sonic horizon. We first outline the physical setup and then investigate details of the proposed experiment. Finally, we will put all the pieces together and present a complete 3D simulation covering all aspects of the outflow scenario.

\subsection{Outflow Scenario \label{analogue_outflow_scenario}}

Consider a BEC reservoir from which the condensate is allowed to flow out. For example, consider the atoms confined between two potential humps in one dimension. If one of these humps is sufficiently weak, the condensate can leak out, with the BEC becoming supersonic and remaining so, avoiding a white hole. We develop this idea of Giovanazzi {\it et al.}~\cite{giovanazzi:horizon} in several ways. First, we simulate the horizon in 3D, exploiting cylindrical symmetry. In particular we do not always assume a quasi-1D situation, and hence study the transverse structure of the horizon. Second, as our discussion in \sref{horizon_temperature} showed the importance of three-body losses, we include them in our simulations. Finally, we evaluate whether the optical piston described in \rref{giovanazzi:horizon} is required in practice.
\begin{figure}
\centering
\epsfig{file={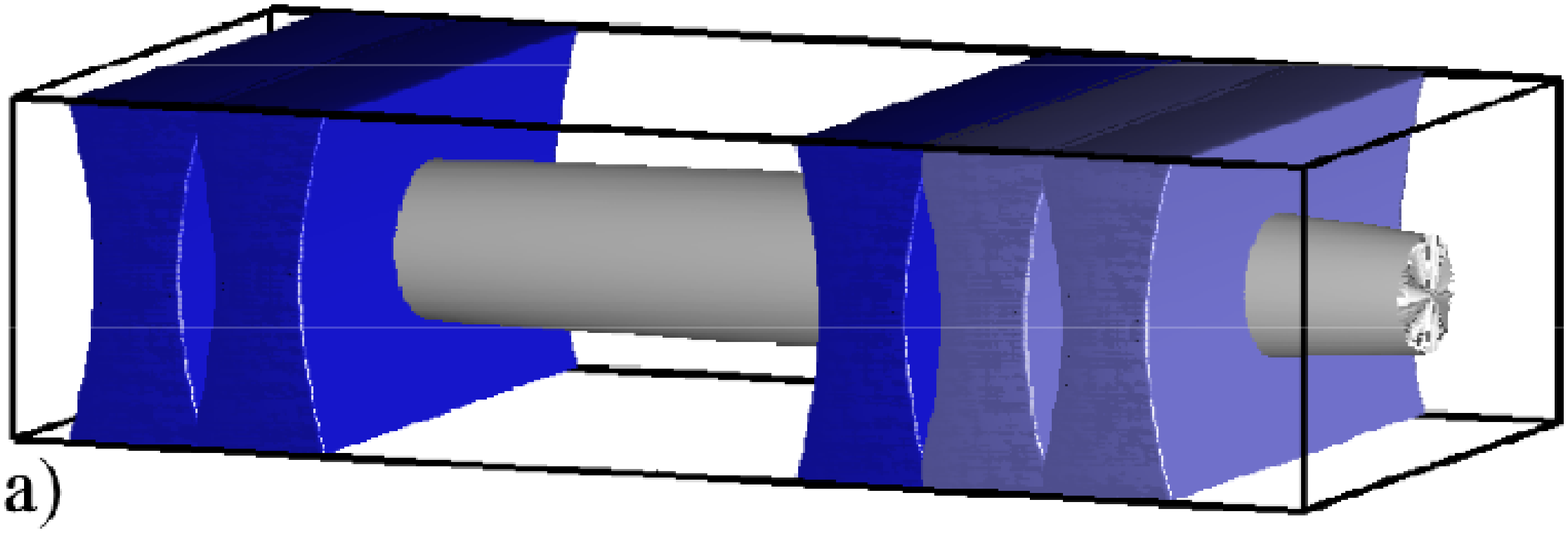},width=0.75\columnwidth} 
\\
\epsfig{file={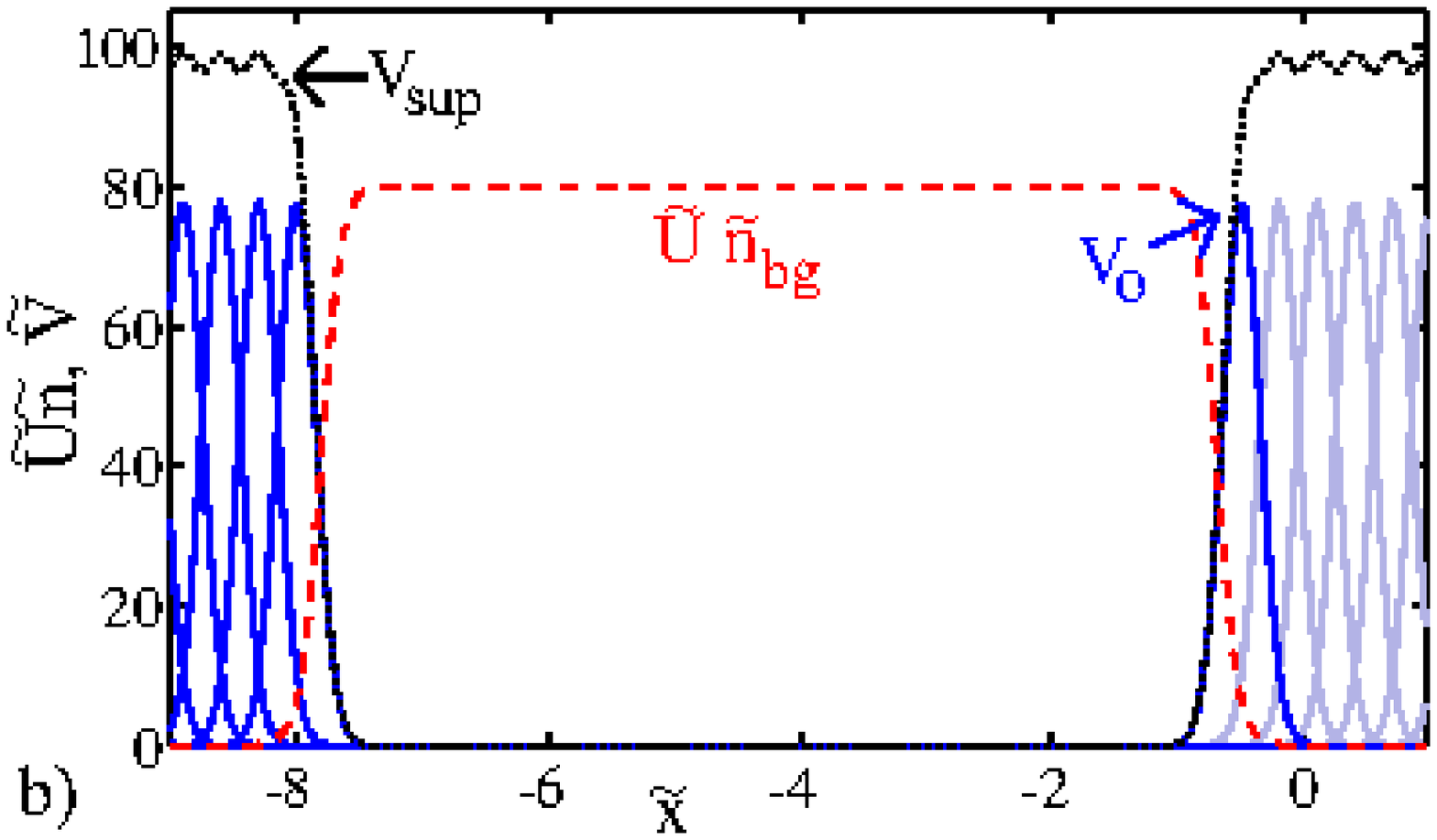},width=0.8\columnwidth} 
\caption{(Color online) (a) Sketch of an elongated condensate (gray), sliced by multiple blue detuned laser sheets (blue).  (b) These sheets (blue) add up to form barriers (black, dotted), which are impenetrable for the condensate. The red, dashed line is the interaction energy  $\tilde{U} \tilde{n}$. The left sheets act as a permanent endcap. To initiate supersonic outflow from this reservoir, the optical potentials sketched in light blue in panels (a) and (b) are suddenly removed. 
\label{3dpiston}}
\end{figure}

A sketch of the outflow scenario is given in \fref{3dpiston}.
The condensate is initially confined in an elongated, cigar shaped trap, sliced by blue detuned laser sheets which act as endcaps. This defines our reservoir. The sheets superimpose to achieve approximately top-hat shaped potentials. The height of this superposition potential 
$\sub{V}{sup}$
is larger than the bulk condensate interaction energy $\tilde{U}\tilde{n}_{bg}$, hence the atoms are confined, see \eref{hydrodyn2}. $\tilde{n}_{bg}$ is the background density within the reservoir, i.e.~where the potential vanishes. To initiate outflow, all but one of the right-hand sheets are suddenly removed. The height of the remaining single sheet $V_{0}$ is less than that required for confinement so that the BEC starts to stream out. The purpose of the superposition method is to avoid a rapid temporal change in the potential within the region where the condensate density is significant, which would disrupt the reservoir. 

In an experiment one could just let the condensate propagate supersonically along the waveguide created by the transverse confinement. However, the numerical treatment requires absorption at the grid edge, for which we employ a simple imaginary potential.  

Taking all these elements together, the potential that we apply for our dynamics is
\begin{align}
\tilde{V}(\dl r,\dl x,\dl t)&=\frac{1}{2}\dl r^{2} + \sum_{n=0}^{N_{l}}V_{G}(V_{0},\dl x_{l}(\dl t)-n\dl d,\dl \sigma)
\CR
&+ V_{G}(V_{0},\dl x_{r}, \dl \sigma) + V_{G}(-i\dl \gamma,\dl x_{oc},\dl \sigma_{oc}) 
\CR
&+\theta(-\dl t)\sum_{n=1}^{N_{r}}V_{G}(V_{0},\dl x_{r}+n\dl d,\dl \sigma),
\CR
V_{G}(W,\dl x_{0},\dl \sigma_{0})&\equiv W\exp{\left(-2 (\dl x-\dl x_{0})^{2}/\dl \sigma_{0}^{2} \right)}.
\label{potentials}
\end{align}
$V_{0}$ denotes the amplitude and $\dl \sigma$ the width of an individual Gaussian sheet. $\dl x_{l}(\dl t)$ is the position of the innermost sheet in the left barrier and $\dl x_{r}$ the same on the right. $\dl d$ is the separation between the sheets. $N_{l/r}$ are the numbers of sheets in the left and right barriers. Usually $\dl x_{l}(\dl t)=\dl x_{l}$ is constant. The absorption is parametrized by $\dl \gamma$, $\dl x_{oc}$ and $\dl \sigma_{oc}$. The potential $V_{G}(V_{0},\dl x_{r},\dl \sigma)$ is distinct from the other members of the right top-hat array, as it is the one that remains after $\dl t=0$. We will refer to this sheet as the ``hump potential''. $\theta(-\dl t)$ denotes the Heaviside step function. For our one dimensional simulations, the potential is given by \eref{potentials} with $\dl r=0$.

Instead of initiating outflow by lowering the barrier on the right, we can also proceed as in \rref{giovanazzi:horizon}. We then use a hump potential just high enough to confine the BEC by itself, $V_{0}=\tilde{\mu}$. Then, simultaneously with the change in potential, we use the left sheets as a moving ``optical piston'': We let $\dl x_{l}=\dl x_{l}(\dl t)=\dl x_{l}(0)+\dl v_{p}\dl t$. The sheets move with the velocity $\dl v_{p}$ to the right and press the condensate out of the reservoir. When we model this, we additionally phase imprint the velocity $\dl v_{p}$ onto the condensate \cite{gajda:imprint}, which makes the outflow smoother.   
 
When we consider three-dimensional situations, we solve the dimensionless variant of \eref{3dgpe} using cylindrical coordinates $\tilde{r}$, $\tilde{x}$, $\varphi$. Besides small details in the shape of the laser sheets, which can be seen in \frefp{3dpiston}{a}, our problem is cylindrically symmetric. We neglect these details and assume independence of the potentials and wave function of $\varphi$. The simulations are thus done on a two dimensional grid. Nonetheless we refer to them as ``3D simulations'', as their results should pertain to the fully three-dimensional scenario due to the approximate symmetry. We find the condensate initial state using imaginary time evolution \cite{garcia:numerics}.  The condensate is assumed to be in the ground state before $\dl t=0$. 

If we consider the effective cross-sectional area $A(x)$ to be constant, the dimensionless analogue Hawking temperature at the hump potential (for $\dl t>0$) can be readily evaluated from Eqs. (\ref{hawkingtemp1}), (\ref{machderiv}) and  (\ref{potentials}). One finds \cite{giovanazzi:horizon}:
\begin{align}
\tilde{T}_{H}&=\frac{1}{2\pi}\sqrt{-\frac{3}{4}\frac{\partial^{2} V}{\partial \dl x^{2}}}=\frac{1}{2\pi}\sqrt{\frac{3V_{0}}{\dl \sigma^{2}}}.
\label{hawkingtemp3}
\end{align}
We know from \sref{horizon_temperature} that Hawking temperatures increase with density. This seems not to be the case in \eref{hawkingtemp3}, however there is
a disguised dependence on $\dl n$ as we are considering the case where $V_{0} \approx \tilde{\mu}=\tilde{U} \tilde{n}_{bg}$. 

\subsection{Speed of Sound at the Horizon\label{sos_horizon}}

To obtain the most general temperature limit in \sref{horizon_temperature}, we have previously derived the speed of sound at the horizon directly from the maximal density allowed by three body losses. In the outflow scenario presented in this section this overestimates the limit, as the density is usually much lower at the horizon than in the bulk condensate. Specifically we can evaluate \eref{hydrodyn2} at the horizon:
\begin{align}
\mu&=\frac{3}{2}m c(x_{h})^{2} +V(x_{h})=\frac{3}{2} m c(x_{h})^{2} +V_{0}
\label{exitdensity1}
\end{align}
In the first step we have used $v=c=\sqrt{U n/m}$ at the horizon and in the second step the fact that the horizon is always established at the peak of the hump potential \cite{giovanazzi:horizon}. By the same equation (\ref{hydrodyn2}) we can estimate the chemical potential as $\mu=U n_{bg}$ since the velocity of the the bulk condensate is small and the potential vanishes there. Solving \eref{exitdensity1} for $c(x_{h})^{2}$ we obtain from \eref{temperaturelimit1}:
\begin{align}
T_{H}&\lesssim \frac{\sqrt{2}\left (U n_{bg} -V_{0}\right)}{3\pi k_{B} D}.
\label{temperaturelimit3}
\end{align}
Hence the strength of the hump potential $V_{0}$ has to be carefully adjusted. It should be large to increase the Hawking temperature \eref{hawkingtemp3} but must be substantially less than the bulk pressure $U n_{bg}$ for the temperature not to violate \eref{temperaturelimit3}. Our numerical results in the following sections are chosen near these limits: We consider bulk densities as large as allowed by losses with a strong hump potential. As a consequence the flow at the horizon will show small deviations from the hydrodynamic regime. We provide the healing length in the background condensate $\dl \xi_{bg}$, the healing length at the horizon $\dl \xi_{h}$ as well as the maximal temperature $\tilde{T}^{*}$ allowed by \eref{temperaturelimit1} (for $D=1$) to relate the horizons to the hydrodynamic conditions. In dimensionless units:
$
\tilde{T}^{*}= \tilde{c}^{2}(x_{h})/(\sqrt{2}\pi)
$.
%

\subsection{Outflow Initiation\label{outflow_initiation}}

In this section we compare the two different methods for initiating the outflow from the reservoir: (i) A sudden change of the right barrier potential from a maximal strength of $\sub{V}{sup}> \tilde{\mu}_{1D}$ to $\sub{V}{0}< \tilde{\mu}_{1D}$. (ii) A sudden change of the right barrier potential from a maximal strength of $\sub{V}{sup}>\tilde{\mu}_{1D}$ to $\sub{V}{0}=\tilde{\mu}_{1D}$ accompanied by beginning to move the left barrier (optical piston) with a velocity $\dl v_{p}=0.01$ to the right and phase imprinting the matching velocity $\dl v_{p}$ onto the condensate. As our investigations here require large reservoir extensions and long simulation times, we resort to a one dimensional model in this and the following subsection, employing the potential $V(\dl r=0,\dl x,\dl t)$ in \eref{potentials}.

\begin{figure}
\centering
\epsfig{file={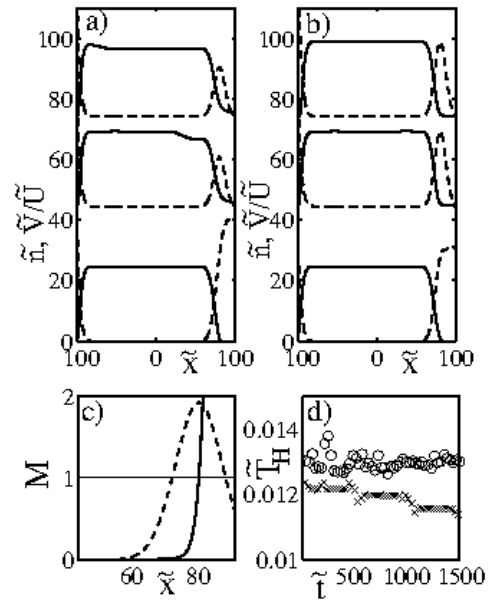},width=0.8\columnwidth} 
\caption{Initial time evolution of the condensate in the reservoir after initiation of the outflow. No three-body loss. (a) Density (solid) and $\tilde{V}(x)/\tilde{U}_{1D}$ (dashed) for outflow with no piston. $\tilde{\mu}_{1D}=0.61$, $\dl \xi_{bg}=0.64$, $\dl \xi_{h}=3.8$, $\dl \sigma=14\simeq22\dl \xi_{bg}$, $V_{0}=0.4$, $\tilde{x}_r=80$, $\tilde{x}_l=-110$, $\dl d=0.5 \dl \sigma$. The time samples are $\tilde{t}$=$0$,  $60$,  $210$, and are shifted up for clarity. The zero of the common axes for $\tilde{n}$ and $\tilde{V}$ for later time samples can be deduced from $\tilde{V}(0)=0$. $\dl T^{*}=0.008$. (b) The same as in (a), but with an optical piston. Here $\tilde{V}_{0}=0.61$, $\dl \xi_{bg}=0.64$, $\dl \xi_{h}=4.66$, $\dl d=\dl \sigma$, all other parameters are as in panel (a). $\dl T^{*}=0.005$. (c) Sample Mach number profile (solid, thick) for the case with piston. The thin black line indicates $M=1$. (d) Comparison of analogue Hawking temperatures with ($\circ$) and without ($\times$) piston.
\label{piston_evaluation_plot}}
\end{figure}

\fref{piston_evaluation_plot} highlights the differences between these methods. Consider panel (a). When the barrier at $\tilde{x}_{r}=80$ is suddenly reduced, the interaction energy $\tilde{U}\tilde{n}$ exceeds $V_{0}$ and the BEC begins to stream out. A wave within the reservoir, resulting from the sudden change of the potential, propagates left, hits the barrier there and reflects. In contrast the initiation with the piston (panel b) shows less perturbation of the reservoir. Panel (c) shows a Mach number profile near the hump potential which represents either case well. Consider the analogue Hawking temperature as a function of time shown in panel (d). For this and similar graphs, we numerically determine the instantaneous location of the horizon and evaluate the dimensionless horizon temperature \eref{dimless_th} there. 

During the timespan covered in panel (d), the speed of sound at the horizon decreases by 40\% if no piston is used. Nonetheless, as predicted by \esref{machderiv}{qdd}, the analogue Hawking temperature remains fairly constant, as the decrease in speed of sound is mostly balanced by a corresponding increase in the Mach number gradient. However, step like decreases in temperature occur. The steps coincide with successive arrivals of the flow perturbation, visible in panel (a), which is repeatedly reflected from the left and right barriers. One has to keep in mind that \esref{machderiv}{qdd} were derived neglecting the quantum pressure term $Q(x)$ and assuming strictly a stationary state.

It is evident in \fref{piston_evaluation_plot} that the density at the horizon is significantly less than the bulk value. The value of the healing length at the horizon $\xi_{h}$ indicates that the horizon in fact slightly violates our demand \eref{hydrodynflow_condition}. This does not affect the conclusions of the present section. The big $\xi_{h}$ has two important consequences: Firstly, the quantum pressure term $Q(x)$ begins to have an effect on the condensate background flow. The effect is small: for \fref{piston_evaluation_plot}, $Q(x_{h})$ initially makes up only 2\% of the energy components in Bernoulli's equation \bref{hydrodyn2}. Secondly, the non-phononic part of Bogoliubov's dispersion relation has an effect on phonon propagation, quantified in \rref{barcelo:diffmetric}. This might have to be tolerated for an outflow experiment for the following reason: The bulk condensate in \fref{piston_evaluation_plot} is as dense as allowed by loss. Thus the only way to increase the density at the horizon, and hence the temperature, is by reducing $V_{0}$ in \eref{temperaturelimit3}. However, if $V_{0}$ is reduced, the bulk perturbation created during initiation becomes much larger.

\subsection{Effect of Three-Body-Loss\label{analogue_outflow_pressure}}

In this section, we show that if inelastic atom collisions are taken into account, the piston is no longer necessary. Without the piston, and without three-body loss, the reservoir density $\tilde{n}_{bg}$ decreases in time, because the reservoir is emptied by the outflow. However, for a large reservoir the fractional atom loss due to the outflow, and the resulting density decrease, will be small. Whether or not a piston is used, 
we expect strong three-body losses in the high density regime, which we are forced to enter to reach high Hawking temperatures. These losses will dominate density reduction. 

\begin{figure}
\centering
\epsfig{file={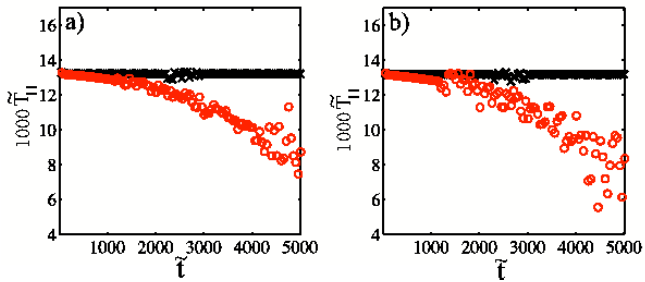},width=\columnwidth} 
\caption{(Color online) Time evolution of analogue Hawking temperature $\tilde{T}_{H}$ for $\tilde{\mu}_{1D}=0.61$, $\dl \xi_{bg}=0.64$, $\dl \xi_{h}=2.8$, $\dl \sigma=14\simeq22\dl \xi_{bg}$, $V_{0}=0.5$. Initially for these scenarios $\dl T^{*}=0.014$. (a) Left end barrier acts as piston with $\tilde{v}_{p}=0.01$. ($\times$-black) $\tilde{K}_{3,1D}=0$. ($\circ$-red) $\tilde{K}_{3,1D}=1.4\times 10^{-7}$. (b) Left end barrier does not move.  ($\times$-black) $\tilde{K}_{3,1D}=0$. ($\circ$-red) $\tilde{K}_{3,1D}=1.4\times 10^{-7}$. $\dl t=5000$ corresponds to $t=117$~ms. The three-body-loss constants are for helium with $\omega_{\perp}=6800\times2\pi$ Hz transverse confinement (\tref{atomicparameters}).
\label{piston_longlosses}
}
\end{figure}

In this section, we employ parameters for helium, assume $\omega_{\perp}=6800 \times 2 \pi$ Hz and a bulk density corresponding to $n=3.6\times 10^{19}$ $\mbox{m}^{-3}$, roughly as in \tref{maxtemperatures1}. We use a large reservoir with $\tilde{x}_{l}=-2270$, $\tilde{x}_{r}=80$, $\dl d=\dl \sigma/2$. Further parameters are given in the caption of \fref{piston_longlosses}. 

We find a significant reduction in the analogue Hawking temperature on the time scale of $50$ ms. While these simulations begin from a state safely in the hydrodynamic regime, within the limits derived in \sref{horizon_temperature}, it does not remain so during the time-evolution: Three-body loss reduces the bulk density and hence also the density at the horizon. As a consequence the local healing length near the horizon approaches the width of the hump potential, hence the flow ceases to obey the hydrodynamic \eref{hydrodyn2}.

We note that the time evolution of the analogue Hawking temperature shows no advantage of the piston setup over free outflow, see \fref{piston_longlosses}. The bulk interaction energy of the initial state condensate $\tilde{U}_{1D}\tilde{n}_{bg}=\dl \mu_{1D}=0.61$ is significantly larger than the exit barrier height $V_{0}=0.5$, giving some room for a reduction of condensate density due to three-body losses without an interruption of outflow.

\subsection{Three-dimensional Horizon Simulations \label{outflow_transverse}}

Finally, we report on a 3D simulation including loss, which shows the feasibility of the proposed experiment. The parameters were chosen to represent the case with the highest analogue Hawking temperature of \tref{maxtemperatures1}: a sodium condensate with $\omega_{\perp}=6800\times2\pi$ Hz ($\sub{a}{osc}=2.5\times10^{-7}$ m), resulting in $\tilde{n}_{3D}=30$, which corresponds to $n_{3D}=1.8\times 10^{21}$ $\mbox{m}^{-3}$. This gives $\tilde{\mu}_{3D}=4.3$, which places the simulation between the Q1D and TTF cases. We choose a scenario that is not in the quasi one-dimensional regime, to show that even if transverse dynamics are possible, a stable horizon is obtained. Further parameters were $V_{0}=2.8$, $\dl \sigma=6.8$. The healing lengths are $\xi_{bg}=0.36$ and $\xi_{h}=1.4$. $T^{*}\times 50=2.8$. Our simulated BEC cloud is about $4$ units ($1$ $\mu$m) in diameter and $70$ units ($17.5$ $\mu$m) long. Following the conclusion of the previous section, we do not use an optical piston.

We find stable outflow with $\tilde{T}_{H}=0.07$ (evaluated at $r=0$), in good agreement with the expectation of $\tilde{T}_{H}=0.068$ from \eref{hawkingtemp3}. For the assumed transverse confinement this equals $23$~nK,  somewhat exceeding the estimate of the achievable temperature given in \tref{maxtemperatures1}. It decays slightly over the timespan of $\tilde{t}=600$ ($14$ ms) of the simulation, largely due to the condensate outflow. Computational limitations required a small spatial domain ($17.5$ $\mu$m), containing only $30000$ atoms. In an experiment the length of the reservoir could be much larger, reducing the temperature reduction due to condensate outflow as shown in \sref{analogue_outflow_pressure}.

A comparison of the atom number evolution with and without three-body loss separates the three-body loss contribution to the reduction of the atom number from that due to outflow. We find that $5\%$ of our initial atoms are lost due to three-body recombination. 

\begin{figure}
\centering
\epsfig{file={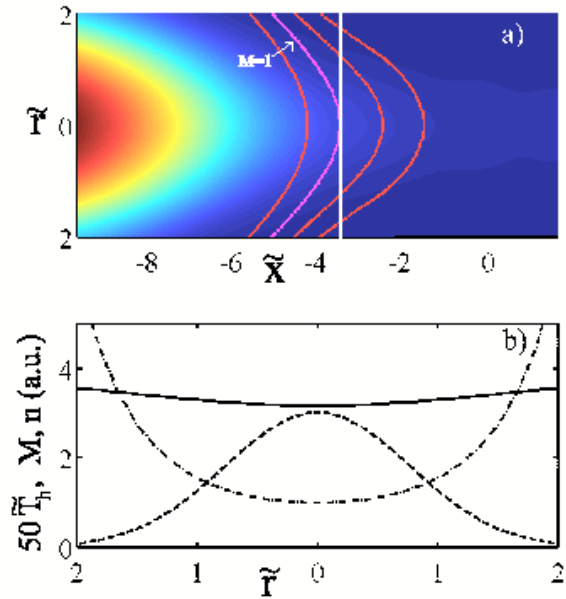},width=0.9\columnwidth} 
\caption{(Color online) Transverse structure of the BEC near the horizon. The full length of the condensate is about 70 units. The hump potential is centered at $\tilde{x}_{r}=-4$ with width $\dl \sigma=6.8$ \cite{footnote:AVinterplay}. The snapshot is taken at $\tilde{t}=525$ ($12$ ms after initation). Other parameters: $\tilde{x}_{l}=-76$, $\dl d=0.7 \dl \sigma$. (a) Colormap of density in $rx$ plane. The density is highest on the left (shaded red) and monotonically decreases to the right (shaded blue). Superimposed are contours of equal Mach number (red) at $M$=$0.5$, $2$, $4$. We marked $M=1$ in violet. (b) Radial structure of Mach number $M$ (dot-dashed), density $n$ (dashed) and proto analogue Hawking temperature $\tilde{T}_{p}$ (solid, see text for definition) at $\tilde{x}=\tilde{x}_{h}=-3.54$. This slice through the condensate is indicated by the white straight line in (a).  
\label{transverse_shape_plots}}
\end{figure}

Our simulation suggests that the Hawking temperatures in \tref{maxtemperatures1} can be realized under realistic conditions, even in a not entirely one-dimensional situation.  \fref{transverse_shape_plots} shows the two dimensional structure of the BEC in the $rx$ plane as well as the radial variation of density, Mach number and the approximate horizon temperature \cite{approx:horizon} $\tilde{T}_{p}=|M'|c/(2\pi)$ for a slice of constant $\tilde{x}=\tilde{x}_{h}$.
This slice does not coincide with the sonic horizon, whose $x$ co-ordinate is a function of $r$. Finding the exact horizon location is difficult. The violet line in \fref{transverse_shape_plots} (a) is the intersection of the $rx$ plane with the surface where $M=1$, which is the boundary of the ergoregion \cite{visser:analogue}. On this surface, the flow velocity equals the speed of sound, but since the flow is not perpendicular to the surface, the velocity's normal component does not. The true horizon slices the $rx$ plane somewhere between the violet and white lines. Due to the spatial proximity of the slice shown in \fref{transverse_shape_plots} and the true horizon, we expect the displayed behaviour of $\tilde{T}_{p}$ to give a very good indication of the transverse temperature structure at the true horizon. At $r=0$ the graph shows the true horizon temperature $\tilde{T}_{h}$.

%

The Mach number increases away from the trap axis, $\dl r=0$. This is largely because the density decreases while the velocity is approximately independent of $r$. $\tilde{T}_{p}$ stays roughly constant on the shown slice. 
This qualitative behavior is found also for simulations with different parameters.

For the shown scenario, the condensate healing length is $\tilde{\xi}=0.36$, which is less than the transverse extension of the BEC, about $\tilde{R} \sim 2$. The transverse structure could thus be resolved by an excitation with wavelength $\tilde{\lambda}\sim1/\tilde{\xi}$, which might affect the excitation's behavior.
In a Q1D scenario, the horizon also has a curved shape, but in that case $\dl \xi \gg \dl R$. Hence the transverse dimension is frozen out for phonons and the radial structure of the horizon should be irrelevant.

The horizon is stable in three dimensions, whether or not the confinement is tight, and for a condensate beyond the Q1D regime its transverse shape is non-trivial.

\acknowledgments
We would like to thank A.~Truscott and H.-C.~N{\"a}gerl for helpful correspondence.
This research was supported by the Australian Research Council under the Centre of Excellence for Quantum-Atom Optics and by an award under the Merit Allocation Scheme of the National Facility of the Australian Partnership for Advanced Computing.


\begin{thebibliography}{35}
\bibitem[{foo({\natexlab{a}})}]{footnote:tcrit}
\bibinfo{note}{The given expression for $\sub{T}{crit}$ is for a homogenous
  ideal gas and shall suffice here. In fact it is modified by interactions and
  confinement, but these modifications are small.}
\bibitem[{app()}]{approx:horizon}
\bibinfo{note}{Unlike $\tilde{T}_{h}$, the approximate temperature
  $\tilde{T}_{p}$ is evaluated at a location near but not at the horizon.}
\bibitem[{foo({\natexlab{b}})}]{footnote:AVinterplay}
\bibinfo{note}{The horizon is not located exactly at $x_{h}=x_{r}$ as one would
  expect from \eref{regularity} for the case $A=const$. This is due to the
  nontrivial transverse structure, allowing an effective, varying $A(x)$.}
\end{thebibliography}
\end{document}